\documentclass[pre, floatfix, letterpaper, twocolumn]{revtex4}
	\usepackage{float}
	\usepackage{amssymb, amsmath}
	\usepackage[colorlinks,citecolor=blue,urlcolor=blue]{hyperref}
	\usepackage{graphicx}
	\newcommand{\InsertFig}[4]
	{\begin{figure}[h!t]
	       \centerline{
	         \includegraphics[width=#4\columnwidth]{./#1}
	       }
	       \caption{{\footnotesize  #2}
	       \label{fig:#3}}
	\end{figure}}
	
	\newcommand{\InsertFigFull}[4]
	{\begin{figure*}[h!t]
	       \centerline{
	         \includegraphics[width=#4\textwidth]{./#1}
	       }
	       \caption{{\footnotesize  #2}
	       \label{fig:#3}}
	\end{figure*}}
	\newcommand{\InsertFigTwo}[5] {
	\begin{figure*}[h!t]
		(a)
         \includegraphics[width=#5\textwidth]{./#1}
         (b)
         \includegraphics[width=#5\textwidth]{./#2}
       \caption{{\footnotesize  #3}
       \label{fig:#4}}
	\end{figure*}}	
	\newcommand{\cN}{{\mathcal{N}}}
	\newcommand{\cW}{{\mathcal{W}}}

	\renewcommand{\O}{\varnothing}		%This is from ambsymb as \O is a text symbol, 
	\newcommand{\kstd}{K}				
	\newcommand{\khn}{\kappa}

	\newcommand{\Eq}[1]{(\ref{eq:#1})}
	\newcommand{\Sec}[1]{\S \ref{sec:#1}}
	\newcommand{\Fig}[1]{Fig.~\ref{fig:#1}}
	
	\newcommand{\App}[1]{Appendix~\ref{app:#1}}
	
\begin{document}
	
\title{Universal Exponent for Transport in Mixed Hamiltonian Dynamics}
%	\title{Universality of the evolution for paradigmatic mixed systems}
	\author{Or Alus}
	\email{oralus@tx.technion.ac.il}
	\author{Shmuel Fishman}
	\email{fishman@physics.technion.ac.il}
	\affiliation{Physics Department\\ Technion-Israel Institute of Technology\\ Haifa 3200, Israel}
	\author{James D. Meiss}
	\email{james.meiss@colorado.edu}
	\affiliation{Department of Applied Mathematics \\ University of Colorado, Boulder, Colorado 80309-0526 USA}

	\date{\today}
	
	\begin{abstract}
We compute universal distributions for the transition probabilities of a Markov model for transport in the mixed phase space of area-preserving maps and verify that the survival probability distribution for trajectories near an infinite island-around-island hierarchy exhibits, on average, a power law decay with exponent $\gamma = 1.57$. This
exponent agrees with that found  from simulations of the H\'enon and Chirikov-Taylor maps. This provides evidence that the Meiss-Ott Markov tree model describes the transport for mixed systems.
	\end{abstract}
	
	%\ams{34C37, 37C29, 37J45, 70H09}
	%\pacs{02.40.-k, 05.45.-a, 45.20.Jj}
	\vspace*{1ex}
	\noindent
	% Keywords:
	
	\maketitle

%%%%%%%%%%%%%
%%%%% Introduction
%%%%%%%%%%%%% 
\section{Introduction}
A typical Hamiltonian system with more than one degree of freedom has a phase space that consists of regular and chaotic regions intertwined in a fractal structure. 
%Dynamical models of realistic systems exhibit a mixed phase space: in some parts of phase %space the dynamics is chaotic and in some parts it is regular. 
In this paper we focus on two-dimensional, area-preserving maps that may arise from Hamiltonian dynamics by Poncar\'e section. For such 2D maps, phase space is partitioned by invariant circles that are absolute barriers, as well as by partial barriers formed from hyperbolic invariant sets such as homoclinic trajectories or cantori \cite{Meiss1992}. Invariant circles can enclose elliptic islands of stability, and these are typically embedded in chaotic zones in a complex, island-around-island structure like that shown in \Fig{Tree}(a). 
	
Two paradigmatic models of this dynamics are Chirikov-Taylor's standard map: 
\begin{equation}\label{eq:StandardMap}
	(\theta',J') = (  \theta+J+
	\kstd\sin\theta, J+\kstd\sin\theta ), 
\end{equation}
and H\'enon's quadratic map:
\begin{equation}\label{eq:HenonMap}
	( x',y') = (-y+2(\khn-x^{2}), x).
\end{equation}
The standard map was introduced by Chirikov and Taylor as a model for interaction between plasmas and electromagnetic radiation \cite{Chirikov1979a}; it models dynamics near any rank-one resonance.
H\'enon proposed his map as a paradigm for local behavior near an elliptic point \cite{Henon1969}; Karney et al.~showed that it is a normal form for a generic saddle-center bifurcation \cite{Karney1982}. 
% and it is also found to be important with relation to Newhouse phenomenon \cite{Gonchenko2010}.
%It models the local structure of a generic elliptic island. 
	
	Invariant structures in chaotic systems can be ``sticky", i.e.,  nearby trajectories may spend a long time in a neighborhood. More precisely, we say a region of phase space is sticky if its survival probability distribution---the probability that a randomly chosen initial condition in the region remains up to time $t$---has a power-law decay \cite{Channon1980}:
	\begin{equation}\label{eq:AlgebraicDecay}
	P_{sur}(t) \sim t^{-\gamma} .
	\end{equation}
As Karney(and many others, subsequently) showed, the outer boundary of an elliptic island is sticky in this sense, though his numerical experiments showed strong fluctuations around what was inferred to be a power-law \cite{Karney1983}.

MacKay, Meiss, and Percival (MMP) proposed that transport through a connected chaotic component bounded by elliptic islands could be described by a Markov model \cite{MacKay1984}.
%{Shenker1982}.  
It was later noted that the ``states" in this model should be connected to form a tree (e.g., \Fig{Tree}), and transport on a Markov tree was studied by Meiss and Ott \cite{Meiss1986a,Meiss1985}. Their model assumed that the tree was self-similar, based on renormalization theory both for the flux through cantori near boundary circles \cite{Greene1986} and for islands-around-islands \cite{Meiss1986}. These calculations gave a power law \Eq{AlgebraicDecay} with $\gamma = 1.97$.
%\cite{\cite{MacKay1983,Meiss1986,Bensimon1987}. 

%%%%%%%%%%%%%
\InsertFigFull{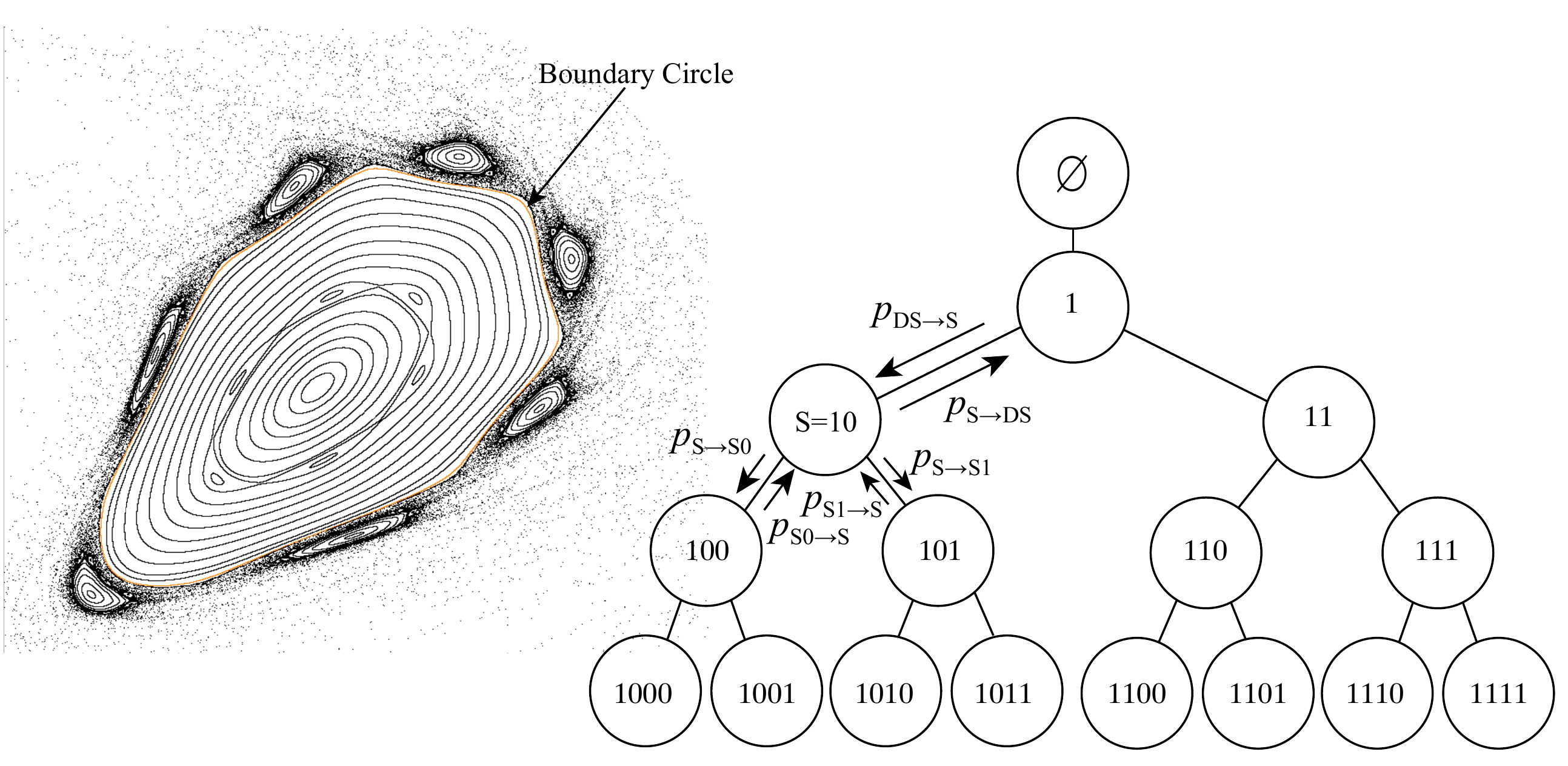}{An elliptic island of \Eq{HenonMap} and a Markov tree. Each node is labeled by the state $S$. Several transition probabilities $p_{S \to S'}$ related to \Eq{FluxRatio} and \Eq{AreaRatio} are also indicated.
The illustrative transition probabilities correspond to state $S = 10$.
}{Tree}{0.8}
%%%%%%%%%%%%%

The self-similar tree model, however, does not explain the commonly observed fluctuations first seen by Karney. Ceder and Agam later showed that if there are uncorrelated fluctuations in the Markov transition rates on the tree, there will be fluctuations in $\gamma$ that decay only slowly as $t \to \infty$ \cite{Ceder2013}. Nevertheless, Cristadoro and Ketzmerick showed that correlated fluctuations in the self-similar scalings of Markov rates will result in a mean decay exponent, $\langle \gamma \rangle$, that depends upon the ensemble;moreover,
if this ensemble is universal for mixed phase spaces, then the mean exponent well be universal as well \cite{Cristadoro2008}. Their numerical simulations of the dynamics of the H\'enon map (without using the Markov tree model) give $\langle \gamma \rangle \approx 1.57$.
	 
In \cite{Ceder2013} and \cite{Cristadoro2008} the ensembles used for the rates were ad-hoc. In this paper, we calculate---for the first time---the transition rates using an ensemble computed from the map \Eq{HenonMap}. 
%Our flux calculations use the action principle of MMP \cite{MacKay1984}. 
We find for the Markov tree %---see \Sec{Markov}---% 
that for the true, dynamical-system-based ensemble, $\langle \gamma \rangle \approx 1.58$.

The outline of the paper is as follows: The Markov tree model is presented in \Sec{Markov}. In particular the results for the survival exponent $\gamma$ are presented there. In \Sec{StandardMap} the survival exponent of the H\'enon map is calculated from the standard map. The results are summarized and discussed in \Sec{conclusions}

%%%%%%%%%%%%%
%%%%% Markov Tree
%%%%%%%%%%%%% 
\section{The Markov Tree Model}\label{sec:Markov}
Consider a phase space with a sticky region formed from an island surrounding an elliptic fixed point such as that depicted in \Fig{Tree}(a). Here we recall the ideas and notation for the Markov tree model for  transport in the connected chaotic component outside such an island  \cite{Meiss1986a}.
	
The fixed point is enclosed by a family of  ``class-zero" invariant circles, the outermost of which is the ``boundary circle"; this circle is one component of the boundary of the chaotic region. Typically there will be a family of broken circles,``cantori", that are outside the boundary circle and that limit upon it \cite{Meiss1992}. The flux of trajectories through these cantori limits to zero at the boundary circle. This gives rise to a set of states in the chaotic region encircling each island, called ``levels", that are bounded by ``partial barriers." In the Meiss-Ott model, these layers correspond to successive rational approximations of the rotation number of the boundary circle.

For the tree of states depicted in \Fig{Tree}(b), the chaotic region ``far" from the sticky region corresponds to the ``root" of the tree, denoted $S =\O$. For calculations of the survival probabiity \Eq{AlgebraicDecay} $\O$ is viewed as absorbing.
The outermost layer surrounding the class-zero boundary circle gives rise to the state denoted by $S = 1$. Successive layers are denoted by adding $1$'s to the state, e.g., $S = 111$ denotes the third layer. 

Within each chaotic layer there will be a largest island chain.
%(typically it will have a rotation number about the elliptic fixed point that is 
%a rational approximation of that of the boundary circle). 
The Meiss-Ott model assumes there is only one such island chain in each layer. Each chain also has a boundary circle, a ``class-one" circle.  The cantori surrounding a class-one circle gives rise to an additional set of chaotic layers. 
The outermost of these class-one layers is denoted $S = 10$. Successive layers approaching the class-one boundary in state $10$ again correspond to adding one's to the state, e.g., $S = 10111\ldots$. 
	
This construction generalizes to each layer: near a class-one boundary there are encircling periodic orbits giving rise to class-two islands, etc. The assumption that there is one island chain in each layer implies that the tree is binary. 
%The address of a state on the binary tree is of the form $S = 10011\ldots$; a one is added when the level increases 
%and a zero when the class increases.

Transport in the connected chaotic region outside all of the boundary circles is thus represented by a sequence of transitions on the tree (levels and classes). If the transport is Markov, it is defined by transition probabilities $p_{S\to S'}$ for each pair of connected states, recall \Fig{Tree}. The probability of such a transition is determined by the flux of trajectories, i.e.,  the area of the turnstile in the cantorus that separates the states  \cite{MacKay1984}. We denote this flux by
$
	\Delta W_{S,S'} = \Delta W_{S',S}; 
$
it is symmetric because the net flux through any region of phase space must be zero. The flux through a cantorus can be computed by the MMP action principle \cite{MacKay1984}. 

The average transit time through a state bounded by such partial barriers is exactly equal to the area of the accessible portion of phase space in the state $S$, $A_S$, divided by the exiting flux \cite{Meiss1997}. If these transit times are long enough that correlations are unimportant, one can assume that the transition probability is
\[
	p_{S\to S'} = \frac{\Delta W_{S,S'}}{A_S}
\]
and that a Markovian approximation is valid.

The only nodes that are connected on the tree are parent-daughter nodes. The daughters of a state $S = s_1s_2s_3 \ldots s_j$ are denoted by concatenation: $S0$ and $S1$. The unique parent of $S$, obtained by deleting the last symbol, is denoted $DS$.
	There are two important transition probabilities, $p_{S\to DS}$ for moving ``up'' from state $S$ to its parent, and $p_{S \to Si}$ for moving ``down'' from a state $S$ to its $i^{th}$ daughter.
It is convenient to categorize the change in transition probabilities from state to state by the two  ratios
\begin{align}
	w_{S}^{(i)}&= \frac{p_{S \to Si}}{p_{S \to DS}} = \frac{\Delta W_{S,Si}}{\Delta W_{S,DS}} ,\label{eq:FluxRatio}\\
	a_{S}^{(i)}&= \frac{p_{S\to Si}}{p_{Si\to S}} = \frac{A_{Si}}{A_{S}}  . \label{eq:AreaRatio}
\end{align}
When the tree is self-similar the ratios \Eq{FluxRatio}-\Eq{AreaRatio} are independent of the state $S$, though they depend on the choice of class, $i=0$, or level, $i = 1$ \cite{Meiss1986a}.  
	
%%%%%%%%%%%%
\InsertFigTwo{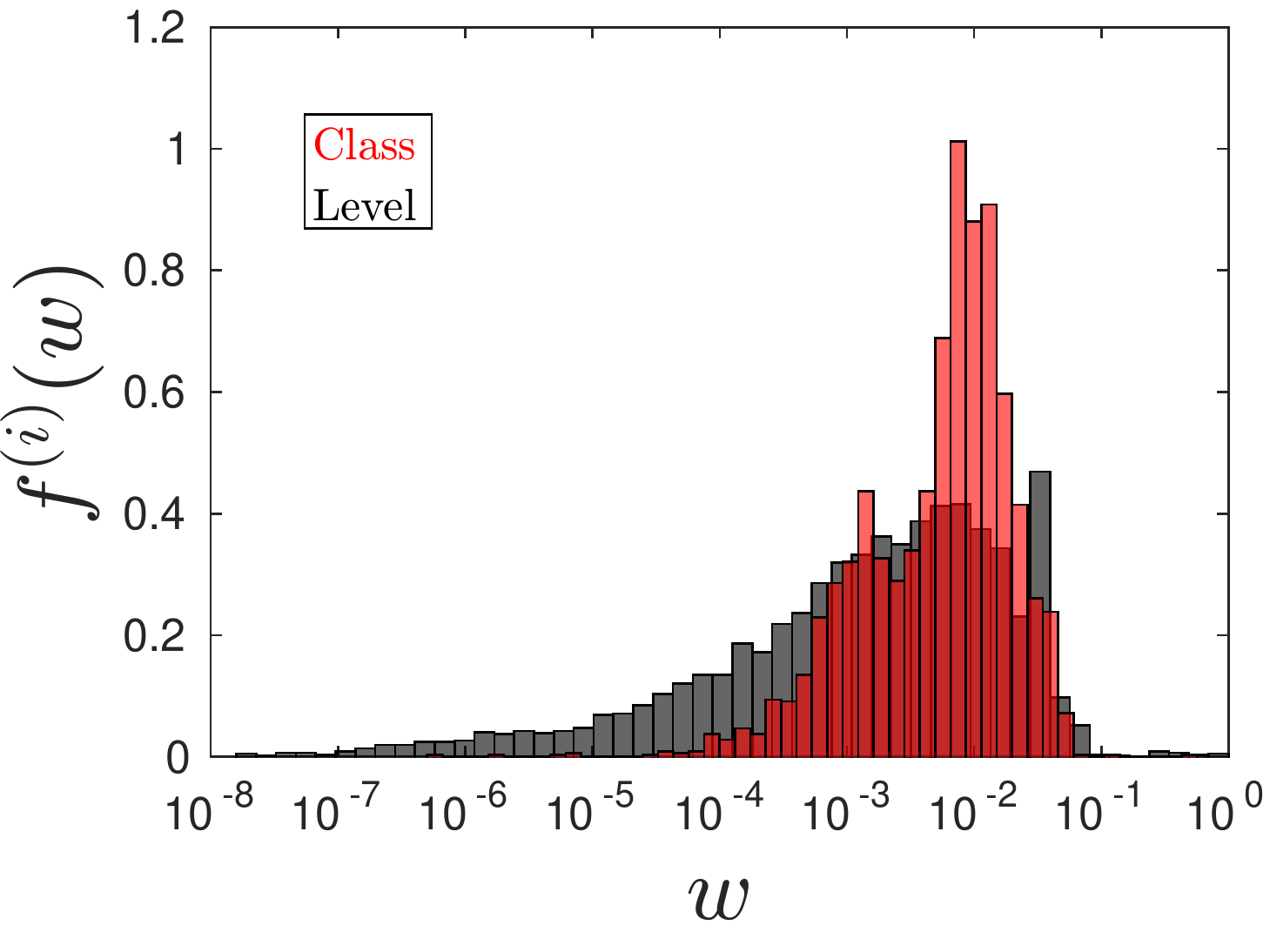}{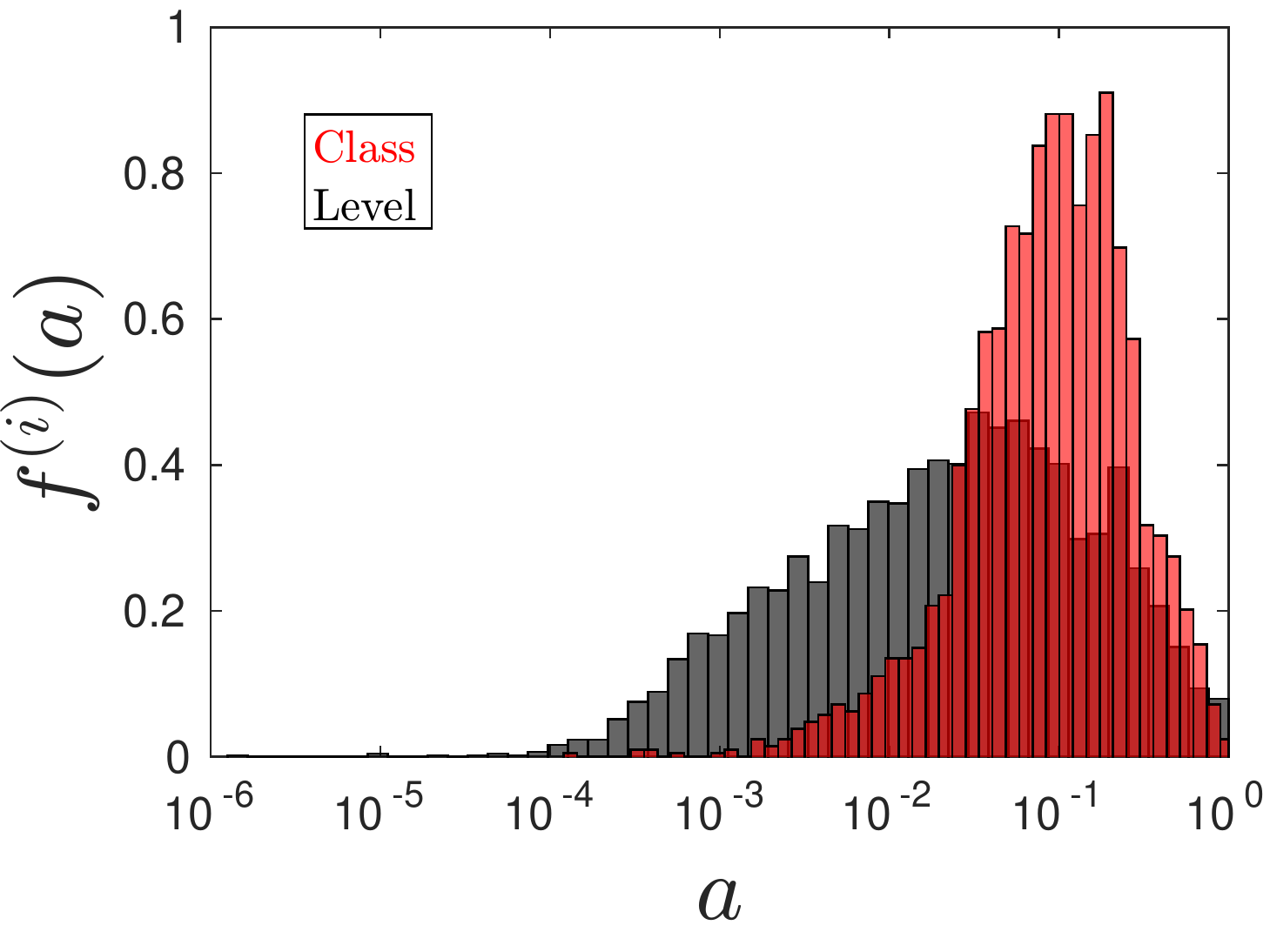}
{(Color online) Distribution densities of (a) area scalings $a$ and (b) flux scalings $w$ for the H\'enon map with $-0.25< \khn <0.75$. Distributions for levels are shown in (dark) gray, and for classes, in red (light gray).
}{Scales}{0.45}
%%%%%%%%%%%%

We previously computed the flux ratios \Eq{FluxRatio} for a number of states and a range of parameter values of the H\'enon map in \cite[Eq.~(22) and Fig.~11]{Alus2014}. The ratios were calculated using fluxes through periodic orbits of \Eq{HenonMap} as a proxy for the cantorus fluxes. The distributions of flux ratios are different for class and level transitions, so we label them as $f^{i}(w)$, see \Fig{Scales}(a). 
Each distribution does not depend systematically on the parameter $\khn$ of \Eq{HenonMap}. In \cite{Alus2014} we compared the $f^{(i)}(w)$ distributions obtained by choosing the parameters uniformly two intervals, $-0.25<\khn<0.25$, and $0.25<\khn<0.75$, (see Fig.~10 and Eq.~(22) there). For the current paper, we repeated this computation using nonuniform $\khn$ distributions (not shown here): the new $f^{(i)}(w)$ do not differ significantly from those in \Fig{Scales}(a). This gives us more confidence in the assertion that there are a ``universal" distributions for class and level flux ratios.

Here we also extend these results by computing the area ratios \Eq{AreaRatio} for a number of orbits of the H\'enon map. Areas were computed for the periodic orbits giving outer rational approximations to the boundary circle rotation numbers up to the states in the fourth generation on the tree; ($S=1001$, $1010$, etc.); 
for \Eq{HenonMap} with $\khn \in [-0.25,0.75]$ \cite{Alus2014}. Under the assumption that the area of the chaotic regions scales as the area of the corresponding regular islands, the area of each island is estimated as that of a polygon defined by a high-period approximation of the boundary circle. The island area ratios \Eq{AreaRatio} were computed up to the third generation on the tree since these require knowing the fourth generation areas.
Figure~\ref{fig:Scales}(b) shows the distributions of the area ratio for class and level scalings. The $f^{(i)}(a)$ again differ significantly for classes and levels, but they still appear to be universal in the sense that they do not depend systematically on $\khn$.
Indeed, \Fig{Uni} shows separate area distributions for the intervals $-0.25<\khn<0.25$ and $0.25<\khn<0.75$---again we see no significant difference to the full distributions in \Fig{Scales}(b).

%%%%%%%%%%%%
\InsertFigTwo{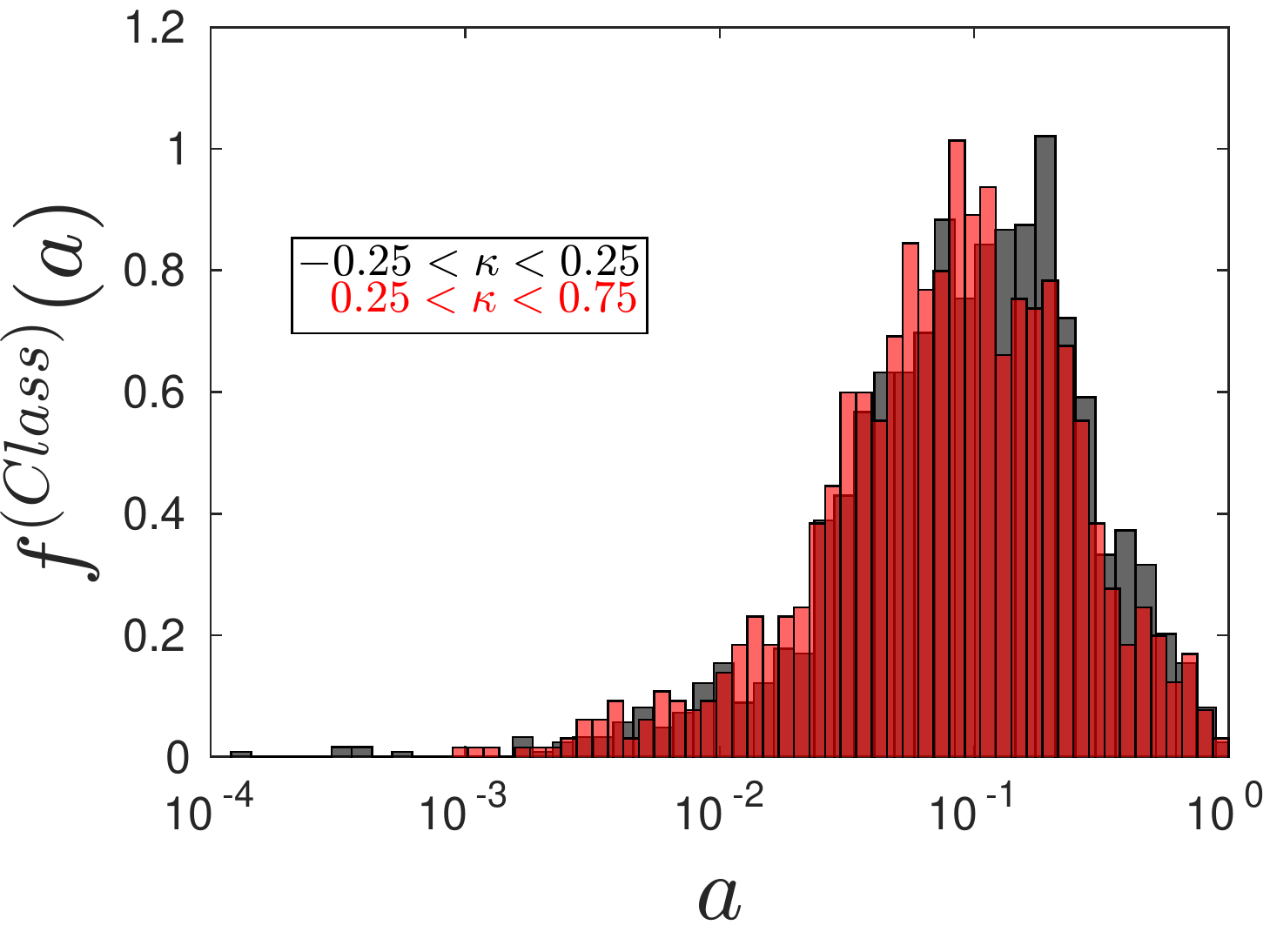}{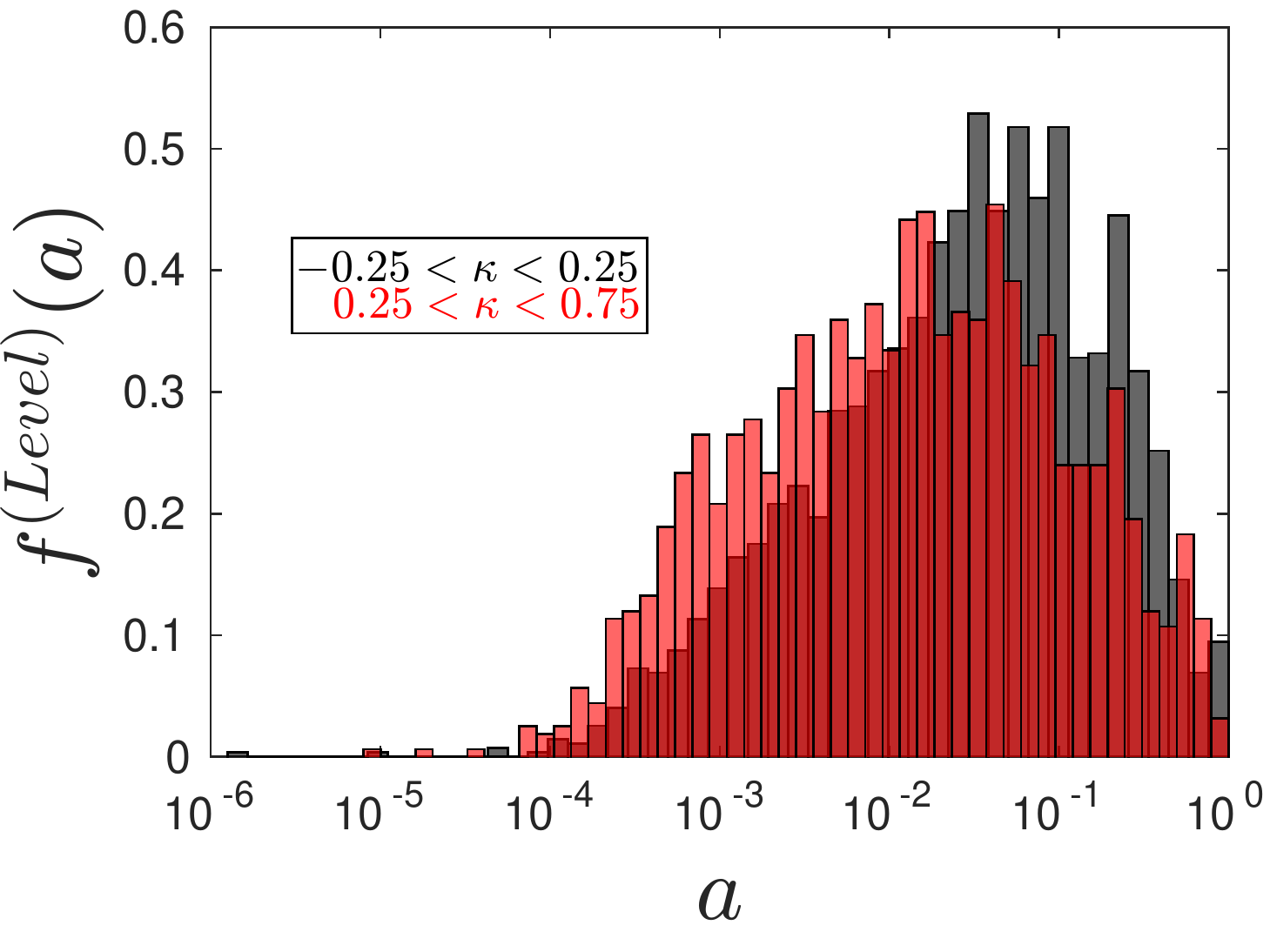}
{(Color online) Distribution densities of (a) class and (b) level area scalings for the H\'enon map with parameters chosen in $-0.25 <\khn < 0.25$, in gray, and in $0.25 <\khn < 0.75$, in red (light gray).
}{Uni}{0.45}
%%%%%%%%%%%%

Finally, we computed the joint probability distributions $f^{(i)}(a,w)$, see \Fig{JointDistributions}. Note that the area and flux ratios exhibit significant correlations, since the probabilities are concentrated on irregular regions in $(a,w)$-space.

Below we use these joint densities to draw values of $a$ and $w$ to give Markov trees with random scalings that correspond, at least according to these first-order statistics, to those of the true map.  That is, we assume that the scaling factors on the different branches of the tree are independent random variables drawn from the empirically computed $f^{(i)}(a,w)$ found from the first three generations of islands and levels for \Eq{HenonMap}. This contrasts with \cite{Meiss1986a} where the ratios for each level and each class branch do not vary with depth on the tree. 

It is important to note that we did not compute the true flux through cantori, nor the true accessible area in any state: we assume that the transition probability rates through the turnstiles of the cantori scale in the same way as those through the numerically computed periodic orbits. Computing the true cantorus flux is considerably more difficult since it must be done using a high-period approximation to the unstable, quasiperiodic cantorus. 
	
In the next subsection, we compute the survival probability exponent $\gamma$ from Monte Carlo simulations on random trees. In \Sec{Master} we compare these results to a master equation approach.

%%%%%%%%%%%%%	 
\InsertFigTwo{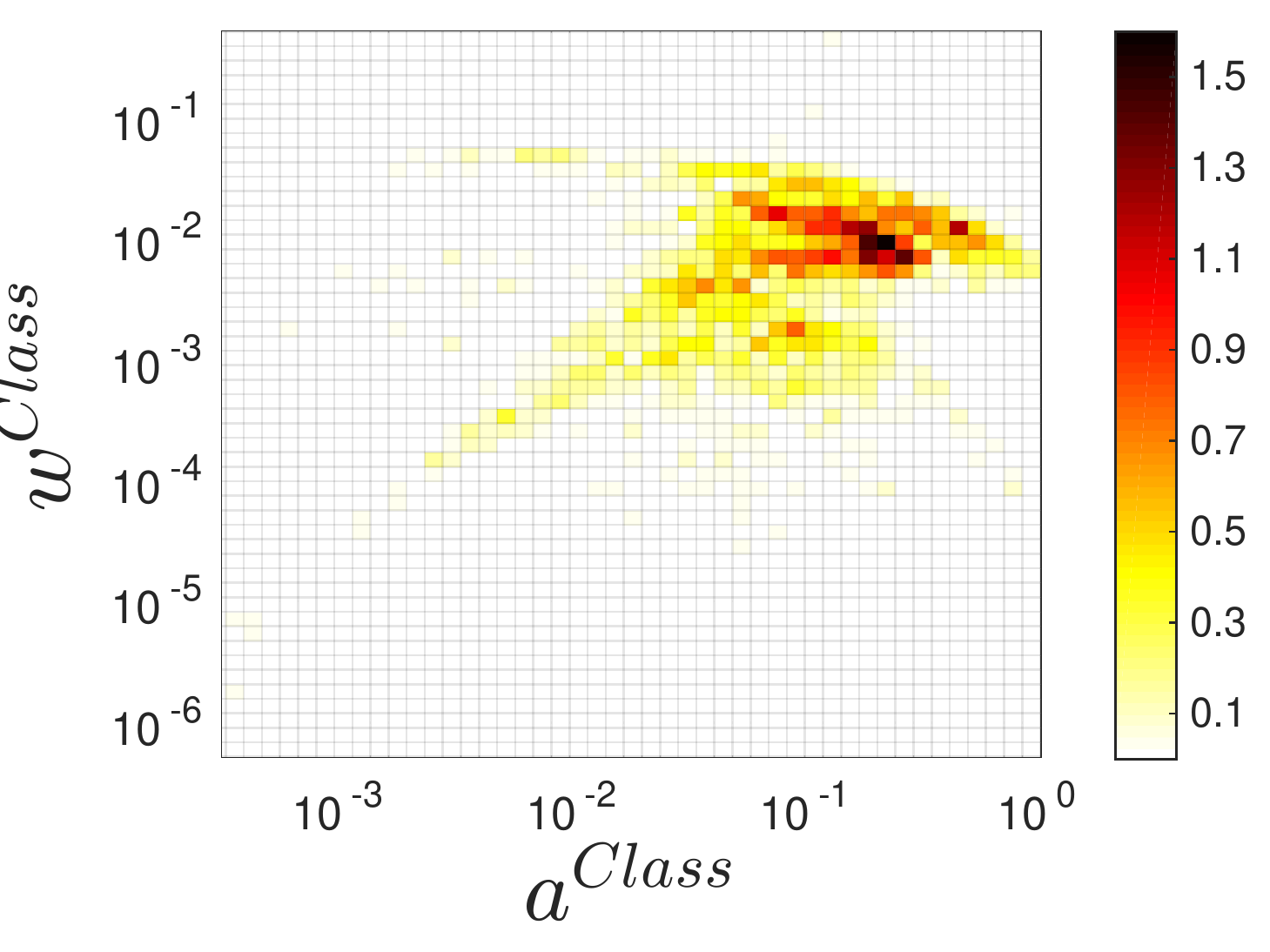}{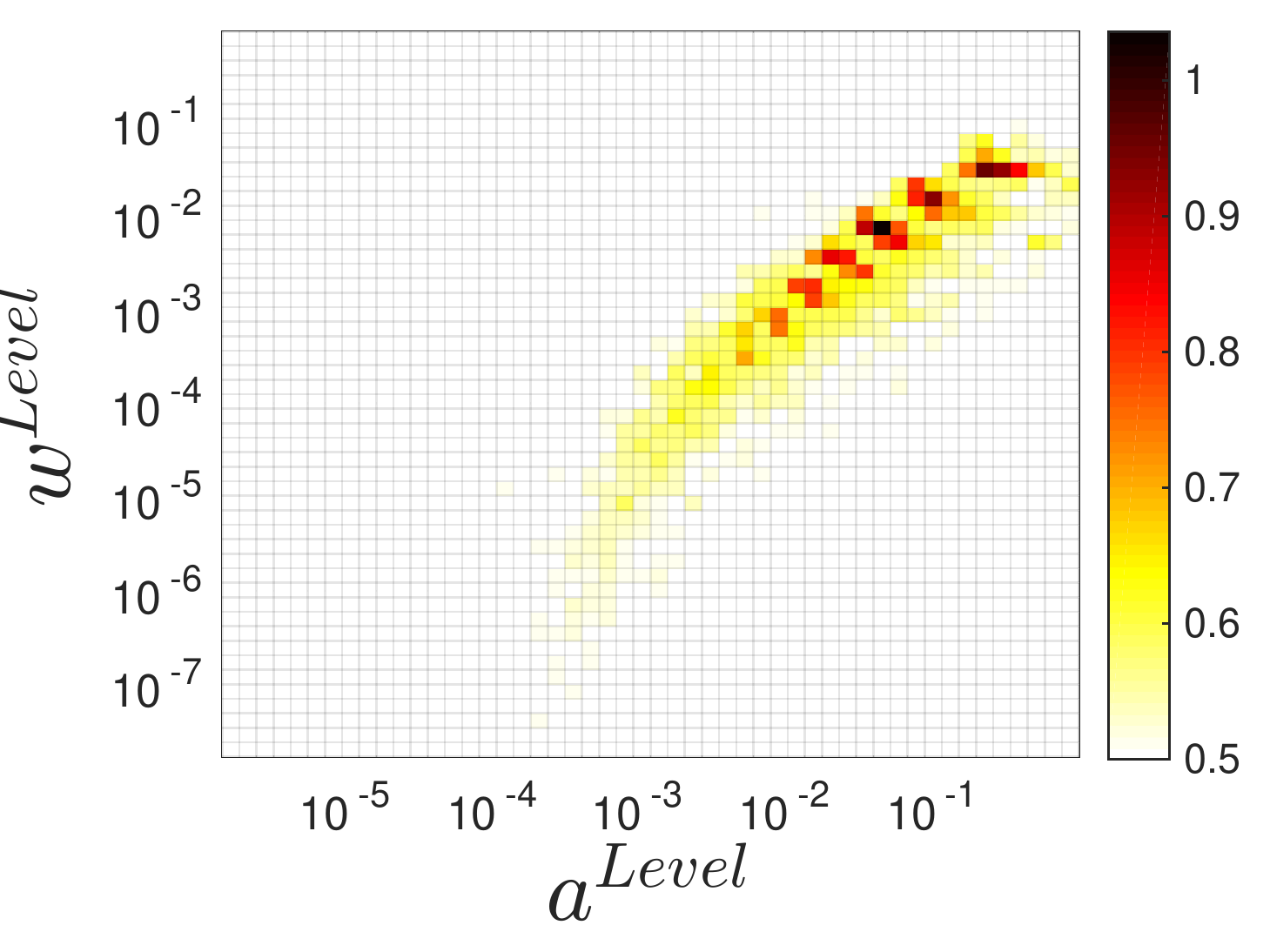}
{(Color online) Histograms of the joint probability distributions for $w$ and $a$. (a) $f^{(Class)}(a,w)$ taken from $2629$ class transitions. (b) $f^{(Level)}(a,w)$ taken from $3608$ level transitions.
}{JointDistributions}{0.45}
%%%%%%%%%%%%%

%%%%%%%%%%%%%
%%%%% Survival Probability
%%%%%%%%%%%%% 
\subsection{Monte Carlo Simulations}\label{sec:survival}
%\textbi{Monte Carlo Simulations---}
For a Markov tree model, the vector of densities at each state on the tree can be denoted by an infinite vector 
$
	 \vec{\rho} = (\rho_\O, \rho_1, \rho_{10}, \rho_{11},\ldots,\rho_S,\ldots)
$,  
where $\rho_S$ is the density at state $S$. If the per-step transition probability is small, transport on the tree is governed by the master equation
\begin{equation}\label{eq:Matrix}
	\frac{d\vec{\rho}}{dt}=\cW\vec{\rho}\,, \quad 
	\cW_{S,S'} = p_{S'\to S}-\delta_{S,S'}\sum_{S''}p_{S\to S''} .
\end{equation}
%or equivalently
%\[
%	\frac{d\rho_S}{dt}=\sum_{S'}\rho_{S'} p_{S'\to S} -\rho_S p_{S\to S'}.
%\]
%Here it is assumed that the density change in one step is very small, so that the difference 
%can be approximated by a time derivative.
The absorbing state, $\O$, is treated by setting $p_{\O\to S} = 0$ for the daughter states $S=1$ or $0$. To
set an overall time-scale we choose $p_{1 \to \O} = 0.1$.  The remaining probabilities are determined by 
the ratios \Eq{FluxRatio}-\Eq{AreaRatio}, which are drawn from the distributions $f^{(i)}(a,w)$ shown in \Fig{JointDistributions}.

Though the tree is infinite, the probabilities for transitions decrease rapidly with level and class, and thus it is reasonable to truncate the tree at a finite number, $B$, of branches or generations.
% such a truncation is inevitable in any real calculation. 
The states in the $B^{th}$ generation are connected only to their parents: only $p_{S \to DS}$ is nonzero. This gives a finite tree with $2^{B}$ states.

To perform the Monte Carlo experiment, we chose $10^8$ particles with initial states drawn from a distribution satisfying detailed balance on the tree \cite{Cristadoro2008}. This is an equilibrium of \Eq{Matrix} when the absorbing state is removed, and since transient behavior is absent, algebraic decay is easier to observe. For such a distribution, the survival probability exponent is $\gamma-1$ \cite{Meiss1997}.
For $B=17$, and averaging over $70$ realizations of the tree we find $\gamma\approx 1.58$. For $B=10$ and $50$ realizations we find $\gamma\approx 1.56$.

%%%%%%%%%%%%%
%%%%% Master Equation
%%%%%%%%%%%%% 
\subsection{The master equation on the tree}\label{sec:Master}
%\textbi{Eigenstates on the tree---}
Here we will compare the Monte Carlo simulations with a direct calculation using eigenvalues $\lambda_n$ and eigenstates $\vec{\rho}_n$ of the $2^{B} \times 2^{B}$ transition matrix $\cW$. The evolution of given initial state $\vec{\rho}(0)$ then becomes
\[
	\vec{\rho}(t) = \sum_{n=1}^{2^B} A_n \vec{\rho}_n e^{-\lambda_n t}, \quad 
	A_n = \langle\vec{\rho_n}^\dagger|\vec{\rho}(0) \rangle
\]
where $\vec{\rho_n}^\dagger$ is the left eigenvector of $\cW$.
The survival probability is
\begin{equation}\label{eq:survivalmatrix}
	P_{sur}(t)=\sum_{S\neq \O} \rho_S(t)=\sum_{S\neq \O} \sum _{n=1}^{2^B} A_n \rho_{n_S}e^{-\lambda_n t}.
\end{equation}
where $\rho_S(t)$ and $\rho_{n_S}$ are the $S^{th}$ component of $\vec{\rho}(t)$ and of $\vec{\rho_n}$, respectively.
To compute \Eq{survivalmatrix}, a reasonable initial condition is 
$
	  \vec{\rho} = (0,1,0,0,0...) .
$ 

As before we use the empirical distributions $f^{(i)}(a,w)$ for the ratios \Eq{FluxRatio}-\Eq{AreaRatio} to generate $\cN = 200$ realizations of a Markov tree. Choosing $B=10$, the decay of $P_{sur}(t)$ appears to be a power law up to $t = 10^{12}$, see \Fig{ExMatrix}. 
The exponent, computed using a least-squares fit from the average $\langle \log_{10}( P_{sur}(t)) \rangle$ for $10^{3.0}\le t \le 10^{12}$ (with equally spaced points on a logarithmic scale) is $ \gamma = 1.50 \pm 0.1$ .

The error in $\gamma$ is estimated from individual realizations: for the upper (lower) bound the product $t^{\gamma_{\pm}} P_{sur}(t)$ exhibits an increasing (decreasing) behavior on a log-log scale for all realizations but one, see \Fig{Error}.
The computed value of $\gamma$ does not change significantly for larger $B$.
%(for example for $B=12$ and $\cN=75$ realizations one finds $\gamma=1.48$). 
The same result is found if one first computes $\gamma$ for each realization, recall \Fig{ExMatrix}, and then average the results. 

In \App{Asymptotic}, we demonstrate how a power law can arise from a sum of infinitely many exponential decays accumulating on $\lambda = 0$.
%%%%%%%%%%%%%	 
\InsertFig{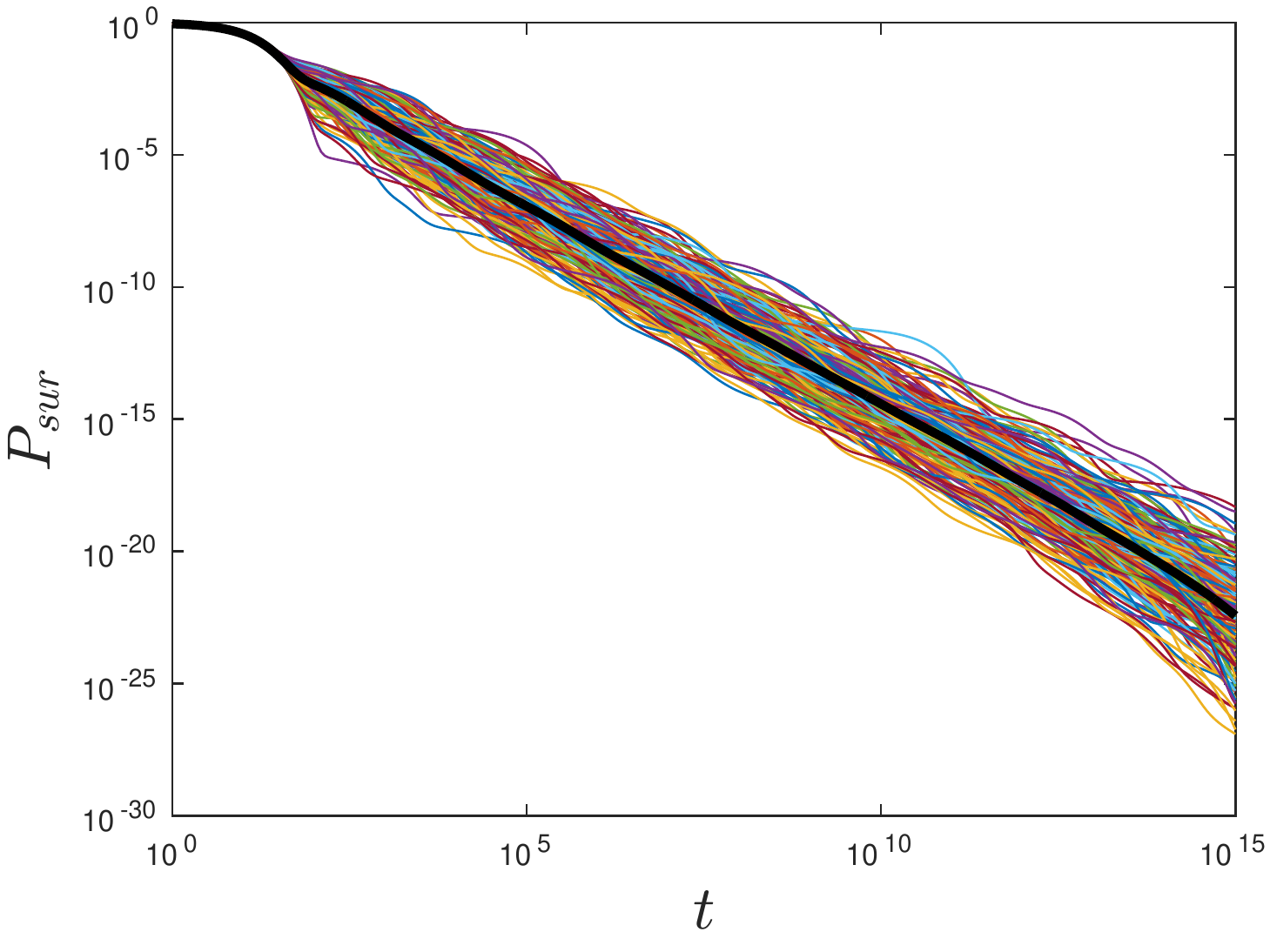}
{Plot of the survival probability $P_{sur}$ vs time for $200$  realizations of the sum \Eq{survivalmatrix}. The heavy line is the average, decaying asymptotically with the slope $\gamma\approx 1.5$ 
}{ExMatrix}{.8}
%%%%%%%%%%%%%
\InsertFigTwo{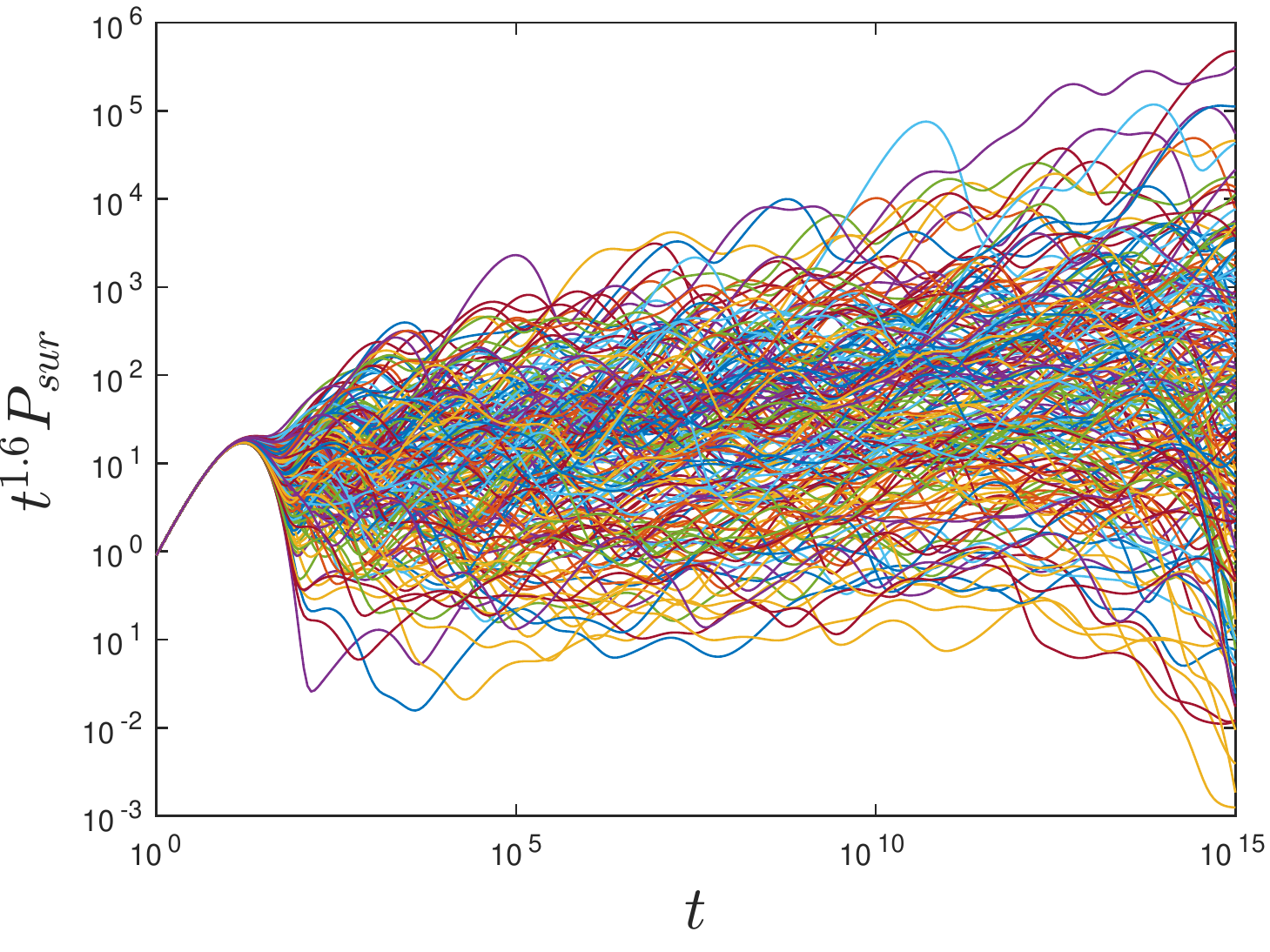}{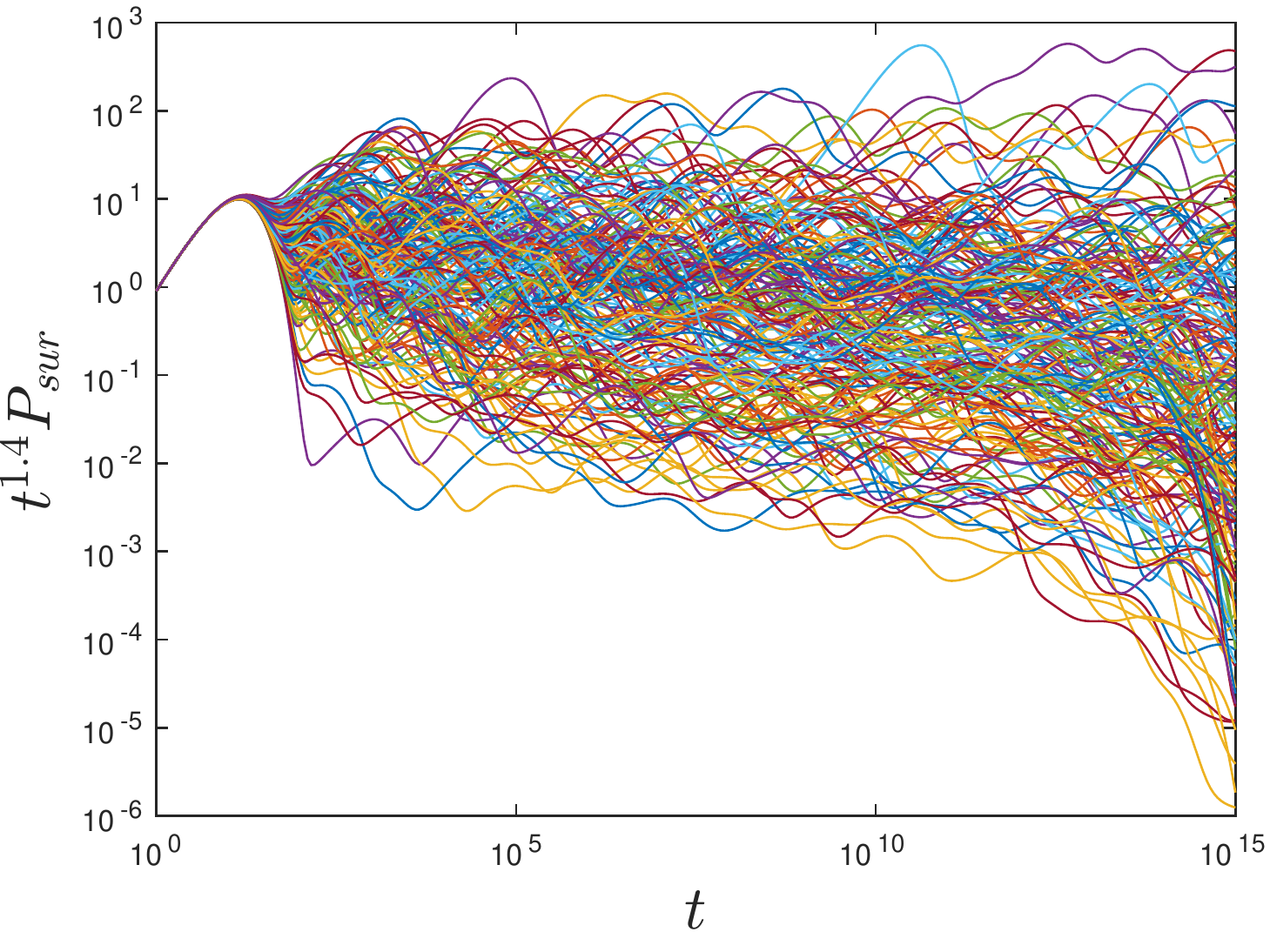}
{(Color online) Plot of the survival probability $t^{\gamma_{\pm}} P_{sur}$  a) $\gamma_+ = 1.6$  and b) $\gamma_{-}=1.4$, vs time for $200$  realizations of the sum \Eq{survivalmatrix}. 
}{Error}{0.45}
%%%%%%%%%%% 

%%%%%%%%%%%%%
%%%%% Accelerator Modes
%%%%%%%%%%%%% 
\section{Stickiness of Accelerator modes}\label{sec:StandardMap}
For large enough values of $\kstd$, the standard map \Eq{StandardMap} exhibits special, 
accelerator orbits for which the momentum increases by a multiple of $2\pi$ each period \cite{Chirikov1979a,Rechester1980,Rechester1981}. These are due to the vertical $2\pi$ periodicity of \Eq{StandardMap}. Indeed taking $J$ mod $2\pi$, accelerator modes are periodic orbits created in saddle-center bifurcations. The simplest of these, at $\kstd = 2\pi n$ for integer $n$, creates two saddle-center pairs; one pair accelerates upward and the other downward. Near a saddle-center bifurcation the local dynamics are modeled by the H\'enon map \cite{Karney1982}. The elliptic points created in these bifurcations remain stable for a small range of $\kstd$, and their neighborhoods are therefore islands like that in \Fig{Tree}(a).

In a regime where there are accelerator islands, the vertical transport in the chaotic component outside the islands is dominated by the stickiness of the islands: trajectories are trapped near the islands with a survival probability \Eq{AlgebraicDecay}. This results in super-diffusion of the momentum \cite{Zumofen1994,Ishizaki1991,Zaslavsky1997,Karney1983}. 

Whenever there is an island with positive acceleration, there is a one with negative acceleration, and the momentum transport can be treated as a random walk between these modes; statistically this is a drunkard's \cite{Karney1983,Ishizaki1991} or a L\'evy \cite{Zumofen1994} walk. 
When trajectories are not stuck, they diffuse in momentum, but the contribution of this gives a negligible contribution to momentum transport. 

%%%%
%\[
%	D(t)= \frac{1}{2t} \langle J(t)^2 \rangle
%\]
%where average is with respect to initial conditions. 

To estimate the exponent $\gamma$, we divide each trajectory into segments that are trapped either near an upward or a downward propagating accelerator island.
Since the upward propagating mode occurs near $\theta = \pi/2$ and the downward one near $\theta = -\pi/2$,
if we take $-\pi \le \theta < \pi$, a transition between the upward and downward motion is corresponds to a change in sign of $\theta$.
When a trajectory is not trapped near an accelerator island, the probability to stay in the same half of the cylinder decays exponentially; therefore the long-time survival probabilities in each half of the cylinder will be dominated by the power-law decay due to the accelerator islands. Computations averaged over 45 parameter values give a survival exponent $\gamma \approx 1.573$.   
These results are in agreement with those of \cite{Cristadoro2008} for \Eq{HenonMap}. For details of the calculation, see  \App{Standard}.

%%%%%%%%%%%%%
%%%%% Conclusions
%%%%%%%%%%%%% 
\section{Results and discussion}\label{sec:conclusions}
%\textbi{Results and discussion---}
We have computed---for the first time---the joint distribution of flux \Eq{FluxRatio} and area \Eq{AreaRatio} ratios for states defined by the island-around-island structure of the H\'enon map \Eq{HenonMap}. 	  
To do this, we assumed that the ratios for cantori scale in the same way as those for periodic orbits. These distributions appear to be universal: they do not depend the parameter $\khn$ of the H\'enon map \Eq{HenonMap} in any systematic way, and this map is the universal local model for an island of an area-preserving map.

Using the Markov tree model, we computed the resulting power-law decay for the survival probability \Eq{AlgebraicDecay} both by Monte Carlo simulations and directly from diagonalization of the transition matrix. 
The mean survival exponent $\gamma$ depends only on the distributions of the scaling ratios and not on the
particular realization of the tree, in agreement with \cite{Cristadoro2008}. Since the scaling distributions are universal, the survival exponent $\gamma$ is universal as well.

Our results are consistent with $\gamma = 1.57$. This is also the value found by direct simulations of the H\'enon map in \cite{Cristadoro2008}. Here, we also found this same exponent for the stickiness of accelerator modes of the standard map.

Thus it appears that the Markov model successfully predicts the algebraic decay exponent observed in simulations. 
The assumptions of the Markov property, the binary structure of the tree, and the use of periodic orbits instead of cantori for the ratios do not negatively impact the results. Therefore the Markov tree is an effective model for the long-time dynamics of transport in area-preserving maps with a mixed phase space.

It remains an open question whether the fluctuations in $\gamma$ are real: namely, do they survive the $t\to \infty$ limit?

%List of results:
%\begin{enumerate}
%	\item The survival exponent was found using accelerator modes of the standard map was found to be $\gamma \approx 
%1.58$ in agreement with Cristadoro and Ketzmerick who found $\gamma=1.57$.
%	\item We found how a power law decay is generated from the combinations of exponential decays for the H\'enon map based Markov tree. (\Sec{Asymptotic})
%	\item Master equation on the Tree (\Sec{Master}) yields $\gamma\approx1.5 \pm 0.1$ for $B=12$ and $\cN= 75$.
%	\item For a simulation of a random walk of particles on the Markov Tree \Sec{survival} one finds $\gamma\approx 
%1.58$
%	\item using integral representation of \cite{Cristadoro2008} assuming random choice of level or class we find $
%\gamma \approx 1.42$
% \end{enumerate}
% 
%What we conclude:
%\begin{enumerate}
%	\item Results 3 and 4 indicate that the Markov tree model , represents the (all?) important physics??
%	\item $\gamma$ is universal (for infinite time ), one value?  no fluctuations? No systematic deviations ?? 
%	\item what about different values of $\gamma$ found in the literature?  Klafter \cite{Zumofen1999comment} 
%	$\kstd = 6.476939$ , $\gamma =1.6$ , Zaslavsky \cite{Zumofen1999comment} $\kstd = 6.476939$, $\gamma =1.58$ Klafter   $\kstd = 6.9115$, $\gamma =1.8$ , Ishizaky \cite{Ishizaki1991} $\kstd = 6.9115 , \gamma =5/3$ 
%\end{enumerate}	  
% 
%Questions:
%\begin{enumerate}
%	\item Still do not know about fixed parameters, $t \to \infty$ as opposed to randomly selected tree.
%	\item Still do not have validation of Markov assumption
%	\item
%\end{enumerate}

%%%%%%%%%%%%%	 
	
{\bf Acknowledgments}: We would like to thank Roland Ketzmerick, Arnd B\"{a}cker, Holger Kantz and Oded Agam for fruitful discussions. OA and SF would like to acknowledge support of Israel Science Foundation (ISF) grants 1028/12 and 931/16, and the US-Israel Binational Science Foundation (BSF) grant number 2010132 and by the Shlomo Kaplansky academic chair. OA acknowledge the support of the Guthwirth foundation excellence fellowship. SF thanks the Kavli Inst.~for Theor.~Phys. for its hospitality, where this research was supported in part by the US National Science Foundation (NSF) under grant NSF PHY11-25915. JDM was partially supported by NSF grant DMS-1211350.

%%%%%%%%%%%%%
%%%%% Appendices
%%%%%%%%%%%%% 
\appendix
\setcounter{figure}{0}
\renewcommand\thefigure{\thesection.\arabic{figure}}    
	  \section{Eigenvalue Asymptotics}\label{app:Asymptotic}
	  A natural question is: how can the master equation \Eq{Matrix} give rise to power-law decay? Indeed, for any finite matrix size, the long-time decay of $P_{sur}(t)$ will be exponential, at the rate of the smallest eigenvalue of $\cW$, say $\lambda_1$. Nevertheless, if $B$ is large enough, then the decay does look like the power-law \Eq{AlgebraicDecay} for a finite time, as we saw in \Fig{ExMatrix}. 

For an infinite chain, a power-law decay over infinite time can occur. For a Markov chain (e.g., keeping only the level transitions), the power law  can arise from a sum of the form 
$P_{sur}(t) \sim \sum\limits_{n} \delta^n e^{-\epsilon^n t}$,
%$P_{sur}(t) \sim \sum\limits_{n} A'_n e^{-\lambda_n t}$,
 implying that the eigenvalues and weights decrease geometrically \cite{Hanson1985}. Inspired by this idea, we note that the long-time behavior of the survival probability depends upon the density of small eigenvalues. Approximating the discrete spectrum by a continuum, then the sum over eigenstates in \Eq{survivalmatrix} becomes an integral,
	  %%%%%%
	  \begin{equation}\label{eq:Integral}
	  	P_{sur}(t) \sim  \int_0^{\lambda_{1}} \sum\limits_{S\neq \O}\rho_{{n(\lambda)}_S} A_{n(\lambda)} e^{-\lambda t} 
	  	\left|\frac{d\lambda}{dn}\right|^{-1} d\lambda .
	  \end{equation}
	  We now suppose that for large $n$, instead of the geometric decay of \cite{Hanson1985}, we have
	  \begin{equation}\label{eq:Asymptotic}
	  	\begin{split}
	  		\lambda_n &\sim n^{-\delta_1}, \\
	  		\sum_{S\neq \O} \rho_{n_S} A_n &\sim n^{-\delta_2} .
	  	\end{split}
	  \end{equation}
To support this hypothesis, we again use the distributions of \Fig{JointDistributions} to compute the eigenvectors  and eigenvalues of $\cW$. The results, shown in \Fig{F_1} for one realization of the tree, show that both of these quantities decrease algebraically with the estimates $\delta_1=5.1$ and $\delta_2=8.8$.
	  Given the asymptotic behavior \Eq{Asymptotic}, \Eq{Integral} becomes
	  \begin{equation}\label{eq:eta}
	  	P_{sur}(t) \sim  \int_0^{\lambda_1} \lambda^{\eta} e^{-\lambda t}  d\lambda \sim t^{-\eta -1}, \quad 
	  	\eta= \tfrac{\delta_2-\delta_1-1}{\delta_1}.
	  \end{equation}
	  For the realization in \Fig{F_1} this gives $\eta = 0.529$.

%%%%%%%%%%%%%	 
\InsertFigTwo{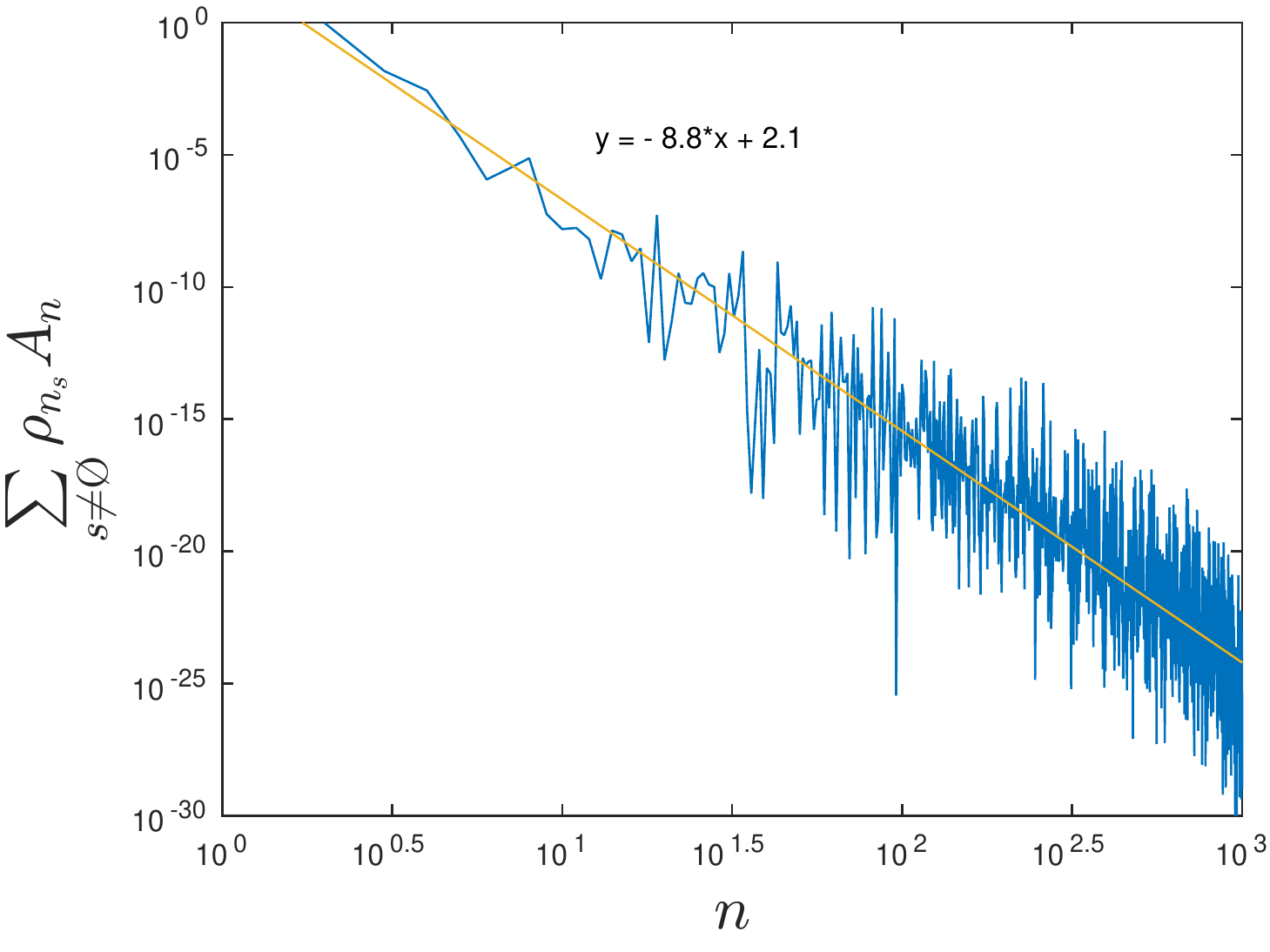}{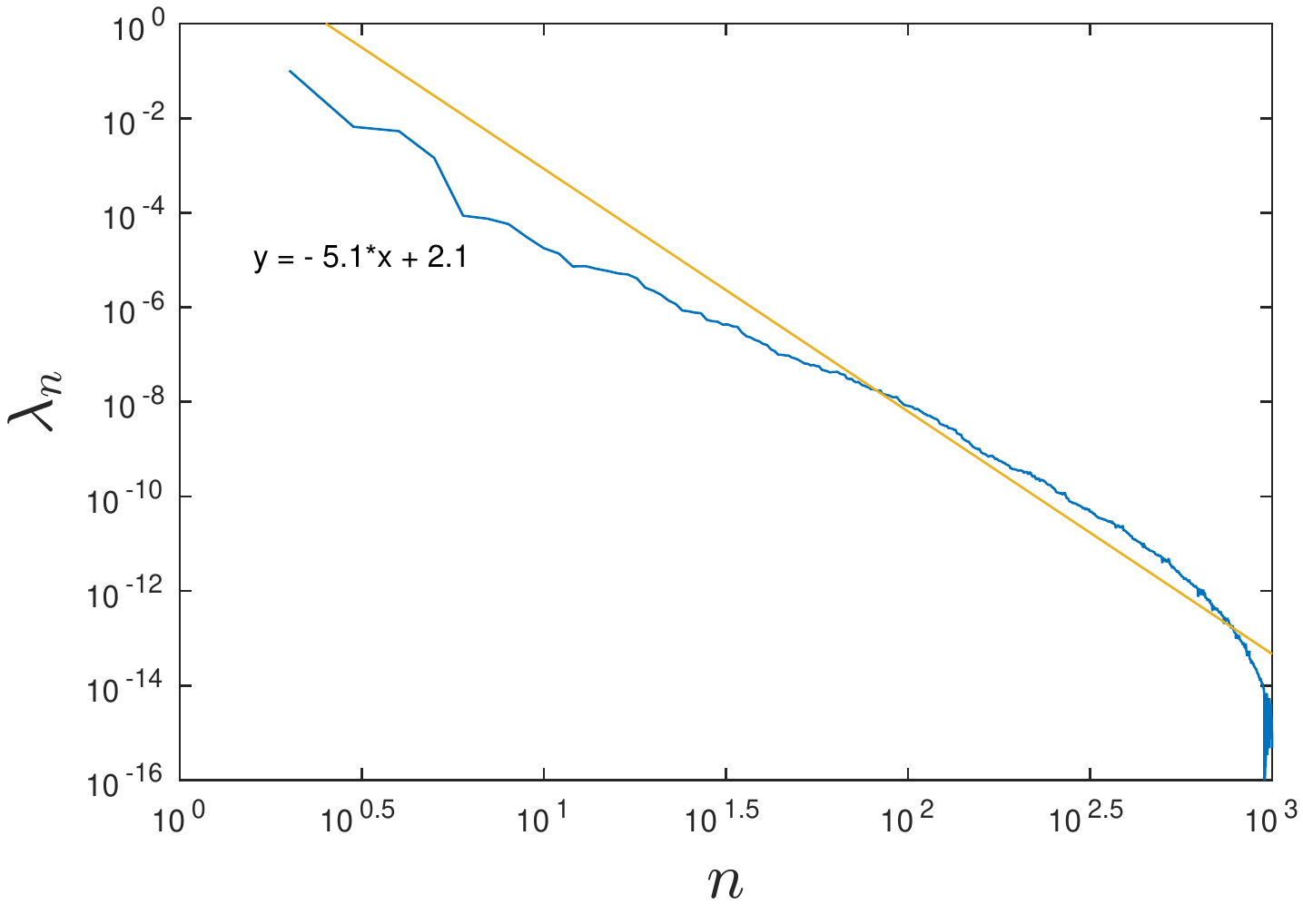}
{Empirical verification of the power laws \Eq{Asymptotic} for a single realization of the matrix $\cW$ in \Fig{ExMatrix}. (a) Plot of $\sum_{S\neq \O} A_n \rho_{n_S}$  vs $n$, and a fit with slope $\delta_2=8.8$. (b) Plot of the eigenvalues $\lambda_n$ of $\cW$ vs $n$, leading to the slope $\delta_1=5.1$. }{F_1}{0.45}   
%%%%%%%%%%%%%
  
We now average the curves, like the one found in \Fig{F_1}, over $\cN = 200$ realizations. 
This assumes that the distributions of $\delta_1$, $\delta_2$, and thus of $\eta$, are narrow so that one can use the average curve to estimate $\delta_1$ and $\delta_2$. Computing the exponents for the averaged curves from points uniformly distributed on a log scale for $n$, we find $\langle\eta\rangle \approx 0.7635\pm 0.16$ where the standard deviation is taken as the error. 
Therefore for the average exponent, \Eq{eta} and \Eq{AlgebraicDecay} imply
	  $
	  \gamma= \langle\eta\rangle+1 = 1.76\pm 0.16. % 1.7635
	  $
	  Alternatively if the fit is done using all values of $n$ (i.e., uniform on the scale of $n$) we obtain 
	  $\langle\eta\rangle\approx 0.490$ and 
	  \begin{equation}{\label{eq: etagamma}}
	  	\gamma=\langle\eta\rangle+1 \approx 1.49.
	  \end{equation}
	  Finally, if we instead calculate slopes for each realization and then compute $\langle\delta_1\rangle$ and $\langle\delta_2\rangle$ and then use \Eq{eta} to find $\langle \eta \rangle$, we find $\gamma \approx 1.70$ and $\gamma \approx 1.48$  for the two fitting methods described above, respectively (uniform in $\log_{10}(n)$, and uniform in $n$). Of the two fits, the latter seems to more appropriately weight the long-time behavior due to the small eigenvalues.

%%%%%%%%%%%%%%
\section{Survival Exponent for the Standard map}\label{app:Standard}
	  
Following Karney in \cite{Karney1983}, we can compute the survival probability from the statistics of the duration of the trapped segments. For example, for a single trajectory of length $T$ with $N$ segments, denote the number of segments of duration $\tau$ by $N_\tau$. 
Then the probability that a segment has length $\tau$ is $p_\tau= N_\tau /N$. 
	  However, to correct for the finite time of the simulations---which over-estimates the probability of observing a short trajectory, Karney showed that one should use   
\[
	  p_\tau=\frac{ N_\tau}{N} \frac{T}{T+1-\tau} .
\]
The cumulative survival probability is then 
\begin{equation}\label{eq:StickyPsur} 
	P_{sur}(t)=\sum\limits_{\tau=t+1}^{T} p_\tau . 
\end{equation}	  
Following the method discussed in the main text to compute $N_\tau$, we computed the $P_{sur}$ for $30$ values of $\kstd$ chosen from equal steps of $0.025$ in the interval $[2\pi,7.8]$ for those cases which had well established super-diffusion; that is, for which no ``singular" islands were present \cite{Zaslavsky1997,Dullin2000}. Singular islands correspond to parameters near the saddle-node, tripling (twistless) and period doubling bifurcations. The omitted parameters also correspond to cases in which the calculation of boundary circles in \cite{Alus2014} failed.
Using a fit with points chosen uniformly in $\log t$ in the interval $10^{2}\le t \le 10^{4}$ gives $\gamma$ values that range over $[1.5, 1.7]$ with an average $\gamma = 1.604$ .  
Adding $15$ more values of $\kstd$ in the interval $[6.4,6.9]$, where again there were no singular islands, leads to 
\[
	  \gamma\approx 1.573 . %1.5735
\]
The results for all $45$ parameter values are shown \Fig{SM}. 
	  
%%%%%%%%%%%%	 
\InsertFig{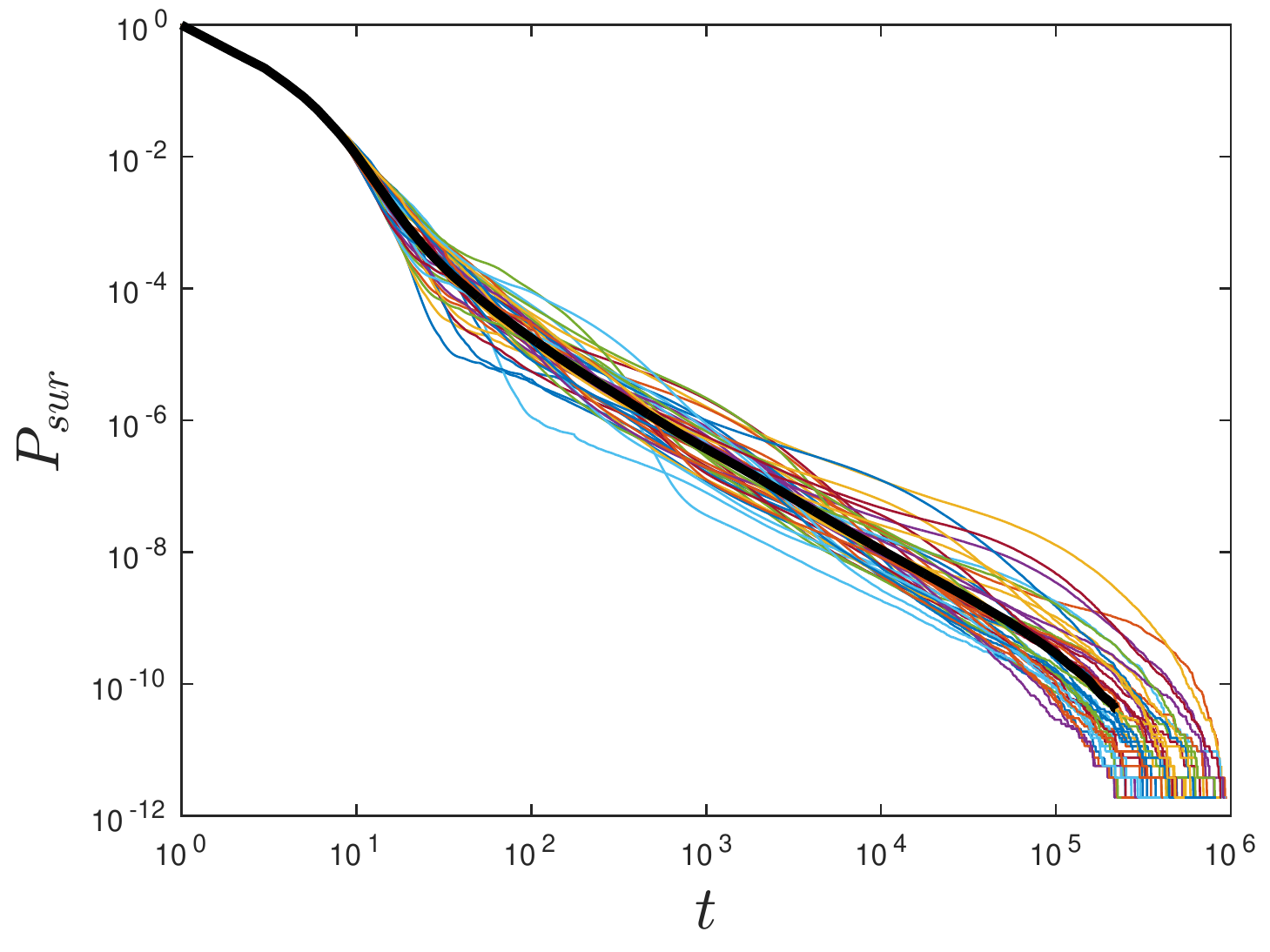}{The survival probability \Eq{StickyPsur} for simulations of \Eq{StandardMap}. The heavy line is the average over $45$ parameter values (see text) resulting in $\gamma\approx 1.573$.}{SM}{.8} 

%%%%%%%%%%%%%%%
%%%%% Bibliography
%%%%%%%%%%%%%%%

\bibliographystyle{apsrev4-1}
\bibliography{Diffusion_sticking}

\end{document}